\documentclass[5p]{elsarticle}
\newcommand{\bc}{\begin{center}}
\newcommand{\ec}{\end{center}}
\newcommand{\benum}{\begin{enumerate}}
\newcommand{\eenum}{\end{enumerate}}
\newcommand{\nn}{\nonumber}
\newcommand{\matl}{\left[ \begin{array}}
\newcommand{\matr}{\end{array} \right]}
\newcommand{\matls}{\left[ \begin{smallmatrix}}
\newcommand{\matrs}{\end{smallmatrix} \right]}

\newcommand{\isdef}{\stackrel{\triangle}{=}}

\newcommand{\rmT}{{\rm T}}

\newcommand{\rmf}{{\rm f}}

\newcommand{\rmq}{{\rm q}}

\newcommand{\BBE}{{\mathbb E}}

\newcommand{\BBR}{{\mathbb R}}

\newcommand{\bbR}{{\mathbb R}}
\newcommand{\T}{\rm T}

\newcommand{\bff}{{ f}}

\newcommand{\bfk}{{ \kappa}}
\newcommand{\bfq}{{ q}}

\usepackage{amsmath}
\usepackage{amsfonts}
\usepackage{amssymb,latexsym}
\usepackage[psamsfonts]{eucal}
\usepackage{amsthm}
\usepackage{graphicx}
\usepackage{epsfig}
\usepackage{psfrag}
\usepackage{subfigure}
\usepackage{pstcol}
\usepackage{epsf}
\usepackage{epstopdf}
\usepackage{amssymb}
\usepackage{amsmath}
\usepackage{setspace}
\usepackage{ulem} 

\newtheorem{theorem}{\indent Theorem}[section]

\newtheorem{proposition}{\indent Proposition}[section]
\newenvironment{prop}{\begin{proposition}$\!\!${\bf }\rm }{\end{proposition}}
\newtheorem{lemma}{\indent Lemma}[section]

\newtheorem{corollary}{\indent Corollary}[section]

\newtheorem{definition}{\indent Definition}[section]

\newtheorem{example}{\indent Example}[section]

\newtheorem{Fact}{\indent Fact}[section]

\makeatletter \@addtoreset{equation}{section}

\makeatother

\title{Estimation of Nonlinear Three-dimensional Constitutive Law \\for DNA Molecules}

\author{Harish J. Palanthandalam-Madapusi\corref{cor1}}
\cortext[cor1]{Corresponding author}
\address{Department of Mechanical and Aerospace Engineering, 263 Link Hall, Syracuse University, Syracuse, NY 13244. USA. Phone: +1 315 443 2107, Fax: +1 315 443 9099, Email: hjpalant@syr.edu}
\author{Sachin Goyal}
\address{Department of Mechanical and Aerospace Engineering, Cornell University, Ithaca, NY 14853, USA, Email: sgoyal@cornell.edu}

\begin{document}
\begin{abstract}
Long length-scale structural deformations of DNA
play a central role in many biological processes including gene
expression. The elastic rod model, which uses a continuum approximation, has emerged as a viable tool to model deformations of DNA molecules. The elastic rod model predictions are
however very sensitive to the constitutive law (material properties)
of the molecule, which in turn, vary along the moleculeÕs
length according to its base-pair sequence. Identification of the nonlinear
sequence-dependent constitutive law from experimental data
and feasible molecular dynamics simulations remains a significant
challenge. In this paper, we develop techniques to use elastic rod model equations in combination with limited experimental measurements or high-fidelity molecular dynamics simulation data to estimate the nonlinear constitutive law governing DNA molecules. We first cast the elastic rod model equations in state-space form and express the effect of the unknown constitutive law as an unknown input to the system. We then develop a two-step technique to estimate the unknown constitutive law. We discuss various generalizations and investigate the robustness of this technique through simulations.

\end{abstract}

\pagestyle{empty}
\maketitle



\section{Introduction}

Designing and engineering DNA molecules to achieve desired biological activity has numerous applications that can lead to advances in disease prevention, diagnosis,
and cure, and is an active area of research. For instance, recombinant DNA technology and gene therapy, which rely heavily on engineered DNA, are revolutionizing the way we treat genetic diseases such as severe combined immunodeficiency (SCID), cystic fibrosis, hemophilia, muscular dystrophy and sickle cell anemia. In addition, designing DNA molecules to achieve desired biological activity is potentially useful in other genetic engineering applications such as engineering algae to produce and synthesize biofuels. 

\begin{figure}[h]
 \centering
 \psfig{height=2.5in,width=3.5in,file=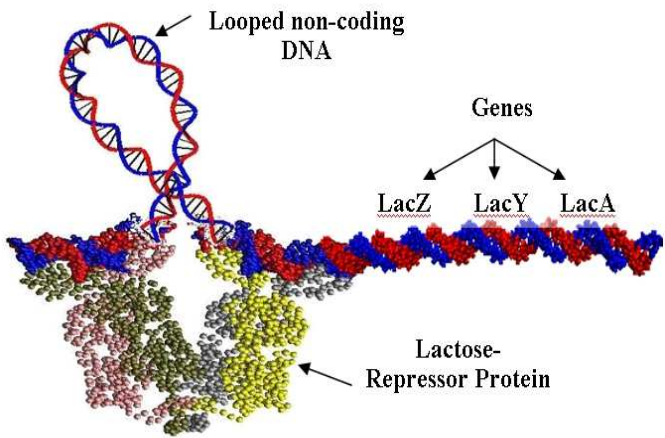}
 \parbox{3.5in}{\caption{\small Schematic of Lac-Operon in bacterium E.coli. This example illustrates the role of structural deformation of the non-coding DNA in regulating genes. The activity of three genes LacZ, LacY and LacA is regulated by looping of adjacent non-coding DNA. The looping is mediated by binding with a V-shaped protein, the Lactose-Repressor.
 \label{fig:Lac_activity}}}
\end{figure}

The biological activities of DNA including gene expression are significantly influenced by its long length-scale structural deformations such as looping \cite{schleif:92a,semsey:05a}, which in turn is tied to its chemical make-up, the base-pair sequence. For example, as shown in Figure \ref{fig:Lac_activity}, activity of the genes in lac-operon in the bacterium {\it E.coli} is governed by the sequence-dependent looping behavior of its non-coding DNA segment adjacent to the genes \cite{goyal:07a}. Thus looping acts as a biological switch to turn on or off the gene expression by restricting access to the transcription initiation site on DNA. In fact, this example has become a paradigm in understanding looping as a common gene-regulatory mechanism.

The success of designed DNA molecules for disease prevention or genetic engineering applications therefore depends not only on the genetic information contained in the DNA but also on the structural deformations of non-coding ``junk DNA.'' Hence there is a need to understand and model the structural deformation of the non-coding DNA segments in order to design DNA molecules that not only have the right genetic information but also undergo desired structural deformations necessary to activate biological mechanism. This is demonstrated by the fact that various designed sequences of non-coding DNA in lac-operon shown in Figure \ref{fig:Lac_activity} affect the gene expression level \cite{goyal:07a,lillian:08a_accepted}.

Moreover, non-coding DNA is a significantly larger part of the genome than the coding part (more than 98 \% in the human genome \cite{collins:04a}). The non-coding part is thus a major consideration for designing DNA and has been mostly ignored thus far. Therefore, understanding the biologically-relevant structural deformations of DNA molecules will greatly accelerate discovery in genetic-disease prevention, diagnosis, cure, and other genetic engineering applications. Both static and dynamic deformations of DNA play a significant role in its biological activity, thus understanding and modeling these deformations and their relationship with base-pair sequences represents a significant challenge \cite{schlick:95a}.

Among the several existing approaches to model structural deformation in DNA molecules
\cite{schlick:02a,goyal:05b,wilson:07b,schlick:95a,olson:96a,balaeff:06a,goyal:07a,goyal:08b,flory:89a,schlick:95, olson:96a, klapper:98,klenin:98a, hsieh:03a}, elastic rod models are based on a continuum approximation, are computationally efficient, and are applicable to long length scales. In this approach, DNA molecules are viewed as continuous filaments. These models have the capability to efficiently represent large nonlinear structural deformations with arbitrary loading and even account for complex interactions \cite{goyal:05b,wilson:07b}. The use of rod models is reasonably well-established in the literature on DNA modeling as reviewed in \cite{schlick:95a} and \cite{olson:96a}. Recent rod models have achieved some promising milestones in describing biologically-relevant deformations of DNA molecules ~\cite{balaeff:06a,goyal:07a,goyal:08b}. However, a key component of these elastic rod models is a constitutive law (material law), which follows from the bond stiffnesses and other atomistic-level interactions and can be approximated by Hooke's law. The constitutive law represents the relationship between the base-pair sequence and the bulk-level elastic properties of DNA molecules, and is largely unknown. A simplistic view of this constitutive law is that it represents the ``springiness'' of the DNA molecule and its relationship to the base-pair sequence.

In macro-scale applications, it is often possible to derive the constitutive law from first-principles, or use experimental measurements to directly determine the constitutive law. However, with DNA molecules these approaches are impractical. Thus the alternative is to use limited experimental measurements to estimate the constitutive law.

Traditionally, the constitutive law is represented by simple elasticity parameters that are tuned by trial and error to match data from specific experiments. For example, average (non-sequence-dependent) bending and torsional stiffness is characterized through single-molecule experiments in \cite{hagerman:88a,strick:96a}. Furthermore, efforts are underway to estimate linear sequence-dependent constitutive law using a massive and systematic collection of molecular dynamics simulations \cite{beveridge:04a,dixit:05a,ponomarev:09a,lankas:09a}. The time and effort involved in trial and error has limited researchers to consider overly simplified models wherein they fit just one or a few parameters \cite{smith:08a}. However, there is no clear consensus on the constitutive law's functional form and how it maps from the base-pair sequence. Recent experimental observations \cite{cloutier:04a,forth:08a} indicate that linear constitutive laws are ineffective in capturing the general structural deformations of DNA molecules \cite{wiggins:05a,smith:08a}. {\it The general form of constitutive law remains largely unknown, and the capability to determine constitutive law of DNA that governs its structure and dynamics is severely limited by the absence of a systematic approach to best leverage the limited data available.}

In this paper, we develop techniques to use elastic rod model equations in combination with limited experimental measurements or high-fidelity molecular dynamics simulation data to estimate the nonlinear constitutive law governing DNA molecules. Moreover, the techniques developed in this paper are also directly applicable to other bio-filaments of medical interest such as collagen fibers, cilia, flagella, and microtubules. We first cast the elastic rod model equations in state-space form and express the effect of the unknown constitutive law as an unknown input to the system. We then develop a two-step technique, in which we first apply simultaneous input reconstruction and state estimation techniques \cite{kitanidis,palanthACC2006,palanthwind2010} to estimate the unmeasured states and the unknown inputs, then use least-squares fitting to estimate the unknown constitutive law. We discuss various generalizations and investigate the robustness of this technique through simulations.

\section{Problem Formulation}

\begin{sloppypar}
We consider experiments in which bio-filaments such as DNA molecules are clamped at one end and loading forces are applied to the other free end. We focus our attention on steady-state deformations and thus ignore transient response. For generality, our subsequent development will be based on bio-filaments. Specific detail relating to DNA molecules will be pointed out periodically.
\end{sloppypar}


Let $s$ be the spatial variable along the length of the bio-filament (rod). Let $f(s) \in \BBR^3$ be the three-dimensional net internal force (tensile and shear) vector, $q(s) \in \BBR^3$ be the net internal moment (bending and twisting) vector, and $\kappa(s) \in \BBR^3$ be the curvature of the rod, respectively, at the location $s$ along the rod. We express all the above-mentioned vectors with respect to a body-fixed reference frame, that is, the basis vectors are attached to each individual cross section of the bio-filament. See \cite{goyal:05b} for details on this reference frame. Furthermore, assuming no external field forces and moments other than boundary forces are applied on the bio-filaments, the elastic rod model describing the motion of the bio-filament \cite{goyal:05b} then simplifies to the following two vector differential equations in terms of the spatial variable $s$
\begin{align}
 \frac{d \bfq}{d s} + \bfk \times \bfq &= \bff \times \hat{ t},\label{eq:ss1} \\
 \frac{d \bff}{d s} + \bfk \times \bff &= 0, \label{eq:ss2}
\end{align}
with the nonlinear constitutive law
\begin{align}
 g(\bfk(s),\bfq(s),\bff(s),s) = 0, \label{eq:const}
\end{align}
where $ g : \BBR^7 \rightarrow \BBR^3$ is a vector function with three components and $\hat { t}$ is the tangent vector to the centerline of the rod.

By setting the origin ($s=0$) to be the free end and the clamped end to be $s=L$, we treat the above problem as an initial value problem by prescribing the loading conditions at the free end. Note that, in general the steady-state deformation of a distributed-parameter system with $s$ as the independent variable is not a causal system and must be treated as a boundary-value problem and not an initial value problem. However, within a cantilever framework, by focussing on the curvature $\bfk(s)$, which is the second derivative of the deformation with respect to $s$, the system is no longer non-causal and can be treated as an initial-value problem.

Thus if the loading conditions at the free end and the constitutive law are known, the rod model equations (\ref{eq:ss1}), (\ref{eq:ss2}) and the constitutive law (\ref{eq:const}) can be solved using a differential algebraic equation (DAE) solver. To simplify the discussion, we assume that the constitutive law can be expressed in the following explicit form with no dependence on $s$
\begin{align}
\bfk(s) = g(\bfq(s),\bff(s)). \label{eq:const0}
\end{align}
Although there is loss of generality in this form and the constitutive law for DNA molecules is known to be base-pair sequence dependent and hence $s$-dependent, it allows us to simplify the discussion and implementation of subsequent methods. We will discuss the more general implicit form (\ref{eq:const}) and $s$ dependence in Section \ref{discussion}. 

Next, to cast the elastic rod model (\ref{eq:ss1}), (\ref{eq:ss2}) in the state-space form with the independent variable as $s$, we define the state vector as $x(s) = \matl{cc} q(s) \\ f(s) \matr\in \BBR^6$, and the input vector as $u(s) = k(s) = \BBR^3$. We then write (\ref{eq:ss1}), (\ref{eq:ss2}) in the state-space form as
\begin{align}
\frac{d}{ds}x(s) & = \frac{d}{ds} \matl{c} \rmq_1(s) \\ \rmq_2(s) \\ \rmq_3(s) \\ \rmf_1(s) \\ \rmf_2(s) \\ \rmf_3(s) \matr \nn \\
 &= \matl{c} \rmq_3(s) \kappa_2(s) - \rmq_2(s) \kappa_3(s) + \rmf_2(s) \\ -\rmq_3(s)\kappa_1(s) + \rmq_1(s)\kappa_3(s) -\rmf_1(s) \\ \rmq_2(s)\kappa_2(s) - \rmq_1(s)\kappa_1(s) \\ \rmf_3(s) \kappa_2(s) - \rmf_2(s) \kappa_3(s) \\ -\rmf_3(s)\kappa_1(s) + \rmf_1(s)\kappa_3(s) \\ \rmf_2(s)\kappa_2(s) - \rmf_1(s)\kappa_1(s)\matr \nn \\
 &= \psi(x(s),u(s)), \label{eq:ss_model}
\end{align}
where $\psi : \BBR^6 \times \BBR^3 \rightarrow \BBR^6.$ A general form of the measurement equation is
\begin{align}
y(s) = h(x,u), \label{eq:meas1}
\end{align}
where $y(s) \in \BBR^l$ represent the measured variables and $h : \BBR^9 \rightarrow \BBR^l$ is the measurement function. The measurements $y$ can either be obtained from experiments on DNA molecules, or from high-fidelity simulations of DNA such as molecular-dynamics simulations.

\begin{figure}
 \hspace{-0.1in}\includegraphics[scale=0.45]{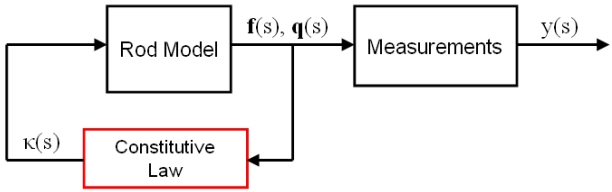}
 \caption{Block diagram representing the relationship between the rod model, constitutive law and experimental measurements. The red block denotes the unknown constitutive law, whose output can be treated as an unknown input to the rod model ({\ref{eq:ss_model}}). \label{fg:block-diagram}}
\end{figure}

We refer to (\ref{eq:ss_model}) as the rod model equations. Note that the unknown constitutive law (\ref{eq:const0}) relating $\bfk(s)$ with $\bfq(s)$ and $\bff(s)$ now becomes
\begin{align}
u(s) = g(x(s)). \label{eq:const1}
\end{align} 
Figure \ref{fg:block-diagram} illustrates the relationship between the rod model, constitutive law and measurement function. Finally, note that if a subset of the states $x(s)$ are measured, then $y(s)$ can be written as
\begin{align}
y(s) = Cx(s),
\end{align}
where $C \in \BBR^{l \times n}$ is a matrix with zeros and ones as its entries.

The problem can then be stated as follows.

{\bf Problem:} Use the measurements $y(s)$ along with known model equations (\ref{eq:ss_model}) to estimate the constitutive law (\ref{eq:const1}).

\section{Constitutive-law Estimation}

First, we note that if measurements of $u(s)$ and $x(s)$ (that is, measurements of $\bfk(s)$,  $\bfq(s),$ and $\bff(s)$) are available, estimating the constitutive law becomes a static nonlinear function estimation problem. 

When either $u(s)$ or $x(s)$ are unknown, estimating the constitutive law is treated as a two-step process. In step 1  we estimate the unknown $u(s)$ or $x(s)$, then in step 2 we use estimates obtained from step 1 to approximate the constitutive law using least-squares fitting. Specifically, we have the following four scenarios:

\begin{enumerate}
\item Measurements of $u(s)$ are available, but $x(s)$ is unknown.

\item Measurements of $u(s)$ and $y(s)$ are available, but the full state $x(s)$ is unknown.

\item Measurements of the full state $x(s)$ are available, but $u(s)$ is unknown.

\item Measurements of $y(s)$ are available, but $u(s)$ and the full state $x(s)$ are unknown.
\end{enumerate}

The above scenarios are summarized in Table \ref{tab:scenarios}.

\begin{table}
\begin{small}
\begin{center}
\begin{tabular}{|l|c|c|l|l}
\hline
& Input $u(s)$ & State $x(s)$ & Estimation Strategy \\[0.7ex] \hline \hline
Function Fit & Measured & Measured & Use (\ref{eq:ls}) \\ \hline
Scenario 1 & Measured & Unknown & Model simulation to estimate $x(s)$ \\ \hline
Scenario 2 & Measured & Partially measured & Kalman filtering to estimate $x(s)$ \\ \hline
Scenario 3 & Unknown & Measured & Input reconstruction to estimate $u(s)$\\ \hline
Scenario 4 & Unknown & Partially Measured & \begin{tabular}{l}Simultaneous state estimation \\ and input reconstruction\end{tabular} \\ \hline
Impossible & Unknown & Unknown & \\ \hline
\end{tabular}
\end{center}
\end{small}
\caption{Various scenarios of available measurements. \label{tab:scenarios}}
\end{table}

In scenarios 1 and 2, since $u(s)$ is known, we can run a numerical simulation using the rod model (\ref{eq:ss_model}) to estimate $x(s)$. In scenario 2 the estimate of $x(s)$ can be further improved by using the measured $y(s)$ with a nonlinear state-estimation algorithm such as the unscented Kalman filter \cite{julier_uhlmann/2004}. In both scenarios, once estimates of $x(s)$ are obtained, we use least-squares fitting using the measured $u(s)$ and estimated $x(s)$ to approximate the unknown constitutive law. An in-depth treatment of scenario 1 and further discussion on scenario 2 are presented in \cite{hinkle2010}.

In the current paper, we focus on scenarios 3 and 4. For these scenarios, we develop a two-step technique for estimating the constitutive law. In both these scenarios, since the input $u(s)$ is unknown, the first step of the two-step technique is an input-reconstruction problem. In scenario 4, in addition to reconstructing inputs, we need to estimate the states. As discussed in the following subsection, the fact that $u(s)$ is not an external input but the effect of an internal feedback nonlinearity as seen in Figure \ref{fg:block-diagram} does not affect the input reconstruction. Once estimates of $u(s)$ and $x(s)$ are available, in step two of the two-step technique, we use least-squares fitting with the estimated $u(s)$ and the measured/estimated $x(s)$ to compute an estimate of the constitutive law. Thus the methodology for estimating the constitutive law can be summarized as the following two-step procedure
\begin{enumerate}
\item[Step 1:] Use input reconstruction (with simultaneous state estimation) to estimate $u(s)$ (and $x(s)$)
\item[Step 2:] Use least-squares fitting with estimates of $u(s)$ and $x(s)$ to approximate the constitutive law
\end{enumerate}

\hrule
\vspace{2.5in}

\hrule

\subsection{Step 1: Input Reconstruction}
\label{sec:IR}

In step 1 of the estimation technique, the objective is to estimate states $x(s)$ and unknown inputs $u(s),$ given measurements of the output $y(s)$. First, we briefly summarize results from \cite{kitanidis,palanthACC2006} for simultaneous input reconstruction and state estimation for a linear discrete system.

\indent Consider the linear discrete state-space system
\begin{eqnarray}
x_{k+1} & = & A_kx_k +  B_ku_k + w_k, \label{eq:mrw1}\\
y_k & = & C_kx_k +  v_k. \label{eq:mrw2}
\end{eqnarray}
where $x_k \in \BBR^n,\,y_k\in\BBR^l,\,u_k\in \BBR^m,\,A_k\in\BBR^{n \times n},\,C_k\in\BBR^{l\times n},$ and $B_k\in \BBR^{n\times m}$. We assume that
$A_k,\,B_k,$ and  $\,C_k$ are known, while $u_k$
is unknown. $w_k \in \BBR^n$ and $v_k \in \BBR^l$ are unknown
Gaussian white noise sequences with known covariances $Q_k$ and
$R_k$ respectively. Without loss of generality, we assume that rank$(B_k)=m$ for some $k$. Finally, we note that $u_k$ is arbitrary and can either be deterministic or stochastic external drivers or be an internal signal such as a nonlinear function of the states.

We consider a filter of the form
\begin{align}\hat{x}_{k+1|k+1}
&= \hat{x}_{k+1|k}+ L_{k+1}(y_{k+1} - C_{k+1}\hat{x}_{k+1|k}),
\label{eq:w3} \\
\hat{x}_{k+1|k} &= A_k \hat{x}_{k|k}. \label{eq:w3a}
\end{align}
Note that since $u_k$ is unknown, the term $B_k u_k$ is absent in (\ref{eq:w3a}).

The state estimation error is
\begin{align}
\varepsilon_k \isdef x_{k+1} - \hat{x}_{k+1|k+1},
\end{align}
and the error covariance matrix is defined as
\begin{align}
P_{k+1|k+1} \isdef \BBE \left[\varepsilon_{k+1}
\varepsilon_{k+1}^{\rmT}\right],
\end{align}
where $\BBE$ is the expected value. The filter is unbiased if and
only if
\begin{eqnarray}
\BBE[x_{k+1} - \hat{x}_{k+1|k+1}] = 0, \label{eq:unbias_basic}
\end{eqnarray}
or consequently
\begin{align}
\BBE[A_k\varepsilon_k & +  B_ku_k +
w_k - L_{k+1}(C_{k+1}A_k\varepsilon_k \nn \\
& + C_{k+1}B_ku_k +
C_{k+1}w_k +v_{k+1})] = 0. \label{eq:unbiasedness1}
\end{align}
Since $u_k$ is arbitrary, (\ref{eq:unbiasedness1}) implies
\begin{align}
(I-L_{k+1}C_{k+1})B_k = 0. \label{eq:constraint1}
\end{align}

Next, we define the cost function $J$ as the trace of the error
covariance matrix
\begin{eqnarray}
J(L_{k+1}) & = &
\mbox{tr}\BBE[\varepsilon_{k+1}\varepsilon_{k+1}^{\rmT}] =
\mbox{tr}P_{k+1|k+1}. \label{eq:cost1} 
\end{eqnarray}

\begin{theorem} \label{thm:complete_filter}
The unbiased minimum-variance gain $L_{k+1}$ in the filter
(\ref{eq:w3}) that minimizes the cost function (\ref{eq:cost1})
subject to the constraint (\ref{eq:constraint1})
is given by
\begin{align}
L_{k+1} = B_k\Pi_k + F_{k+1}\tilde{R}_{k+1}^{-1}(I-V_{k+1}\Pi_k),
\label{eq:finalL}
\end{align}
where 
\begin{align}
\Pi_k & \isdef
(V_{k+1}^\rmT\tilde{R}_{k+1}^{-1}V_{k+1})^{-1}V_{k+1}^\rmT
\tilde{R}_{k+1}^{-1},  \label{eq:k11} \\
\tilde{R}_{k+1}& \isdef 
C_{k+1}P_{k+1|k}C_{k+1}^\rmT + R_{k+1}, \label{eq:defRt}
\\
F_{k+1} & \isdef P_{k+1|k}C_{k+1}^\rmT , \label{eq:defF}\\
V_{k+1} & \isdef C_{k+1}B_{k}. \label{eq:k9}
\end{align}
Furthermore, the covariance update equation is
\begin{align}
P_{k+1|k+1} & = P_{k+1|k} -
F_{k+1}\tilde{R}_{k+1}^{-1}F_{k+1}^\rmT
+ \nn \\
 &(B_k-F_{k+1}\tilde{R}_{k+1}^{-1}V_{k+1})(V_{k+1}^\rmT\tilde{R}_{k+1}^{-1}V_{k+1})^{-1}
 \nn \\
& \times(B_k-F_{k+1}\tilde{R}_{k+1}^{-1}V_{k+1})^\rmT, \label{eq:inv1} \\
P_{k+1|k} &\isdef  A_kP_{k|k}A_k^\rmT + Q_k. \label{eq:defPk}
\end{align}
\end{theorem}

It is straightforward to check that $L_{k+1}$ given by
(\ref{eq:finalL}) satisfies the constraint (\ref{eq:constraint1}). Furthermore, in the absence of unknown inputs, the traditional Kalman filter gain is obtained by setting $B_k = 0$ in the optimal filter gain $L_{k+1}$ given by
(\ref{eq:finalL}).

So far, we discussed unbiased estimation of the state $x_k$ in
the presence of arbitrary unknown inputs $u_k$. Next, we discuss how
the unknown inputs $u_k$ are estimated, using the unbiased estimates
$\hat{x}_{k|k}$ of the states $x_k$.

\begin{prop} \label{prop_d_estimate}
Suppose that $\hat{x}_{k|k}$ is an unbiased estimate of the states
$x_k$ of (\ref{eq:mrw1}). Then
\begin{align}
\hat{u}_k  = B_k^\dag L_{k+1}(y_{k+1} - C_{k+1}\hat{x}_{k+1|k}), \label{eq:d_estimate}
\end{align}
is an unbiased estimate of $u_k$. \label{prop:d_estimate}
\end{prop}
{\bf Proof.} Since $l \geq p$, we can define $\hat{u}_k$ as
\begin{align}
\hat{u}_k  = B_k^\dag L_{k+1}(y_{k+1} - C_{k+1}\hat{x}_{k+1|k}), \label{eq:tempd}
\end{align}
where $\dag$ denotes the Moore-Penrose generalized inverse. Next, we
use (\ref{eq:w3}) and (\ref{eq:tempd}) to get
\begin{eqnarray}
\hat{u}_k \hspace{-0.1in}& = & B_k^\dag (\hat{x}_{k+1|k+1} - \hat{x}_{k+1|k}) \nn \\
  & = & B_k^\dag (x_{k+1} + \varepsilon_{k+1} - A_k \hat{x}_{k|k}) \nn \\
  & = & B_k^\dag (x_{k+1} - A_k x_k + \varepsilon_{k+1} - A_k \varepsilon_k)\nn \\
  & = & B_k^\dag (B_k u_k + w_k + \varepsilon_{k+1} - A_k \varepsilon_k). \label{eq:d_est_1}
\end{eqnarray}
Further, taking expected value on both sides of (\ref{eq:d_est_1}),
yields
\begin{eqnarray}
\BBE [\hat{u}_k] & = & \BBE [B_k^\dag (B_k u_k + w_k +
\varepsilon_{k+1} - A_k \varepsilon_k)], \nn \\\label{eq:d_est_2}
\end{eqnarray}
Finally, noting that $\BBE[\varepsilon_k] = 0$ and the fact that
$w_k$ is zero-mean, we get
\begin{align*}
\BBE [\hat{u}_k] = B_k^\dag B_k \BBE[u_k] = \BBE [u_k].
\tag*{$\square$}
\end{align*}

\begin{sloppypar}
Thus by using Theorem \ref{thm:complete_filter} and Proposition \ref{prop_d_estimate}, we can estimate states and simultaneously reconstruct unknown inputs. The above development focusses on the linear case. The unscented unbiased minimum-variance (UUMV) filter, a nonlinear extension of the unbiased minimum-variance filter that uses the same expression for the gain matrix (\ref{eq:finalL}) and the unscented transform to compute the error covariance matrix $P_{k+1|k+1}$ is discussed in the appendix.
\end{sloppypar}

It is worthwhile to note that since there exists a $k$ such that rank$(B_k) = m,$ it follows from (\ref{eq:constraint1}) that there exists $k$ such that rank$(C_{k+1}B_k) = m$ and therefore $l \geq m$. Finally, since the above development is in discrete space, we use Euler discretization with a small step size to discretize the state-space equations (\ref{eq:ss_model}).


\subsection{Step 2: Least-squares Function Approximation}

Once estimates of both states $x(s)$ and inputs $u(s)$ in (\ref{eq:ss_model}) are obtained, in step 2, least-squares functions approximation tools are used to estimate the constitutive law. In this case, a variety of techniques can be applied to estimate the constitutive law. To use a standard least-squares function approximation technique, we assume that $g$ can be expressed as a basis-function expansion
\begin{align}
u_j(s) = \sum_{i=1}^p \omega_{i,j}\phi_i(x(s)), \quad j = 1,\,2,\,3, \label{eq:ls_fit}
\end{align}
where $\phi_i : \BBR^6 \rightarrow \BBR$ are the basis functions, $\omega_{i,j}$ are the unknown coefficient of the basis function expansion, and $p$ is the number of basis functions chosen. Since $u(s)$ and $x(s)$ are known, and $\phi_i$ are user-chosen, the unknown coefficients $\omega_{i,j}$ are then determined by standard least-squares fitting. The least-squares solution for the unknown coefficients is
\begin{align}
\hat \Omega =  U\Phi^\dag, \label{eq:ls}
\end{align}
where $\Omega = \matl{ccc} \omega_1 & \cdots & \omega_p \matr$ is the coefficient vector, $U = \matl{ccc} u(0) & \cdots & u(L) \matr$ and
\begin{align}
\Phi = \matl{ccc} \phi_1(x(0)) & \cdots & \phi_1 (x(L)) \\ \vdots & \ddots & \vdots \\ \phi_p(x(0)) & \cdots & \phi_p (x(L)) \matr.
\end{align}

\section{Decoupled One-dimensional Problem}

\label{oneDprob}

We start with the simplest case in scenario 3 considered in \cite{law}.

In this section, we first make the following assumptions.
\begin{enumerate}
\item[A1] The material behavior of DNA molecules is decoupled in the principal directions of bending and torsion and is internal-force independent.
\item[A2] The measured quantities are available for all values of $s$.
\item[A3] Measurements are noise-free.
\end{enumerate}

These assumptions will be relaxed in subsequent sections.

In view of assumption A1, by choosing the body-fixed frame along these principal directions, the vector constitutive law (\ref{eq:const1}) is decoupled into the following scalar constitutive law equations
\begin{align}
u_1(s) & = g_1(x_1(s)), \label{eq:mat_law1}\\
u_2(s) & = g_2(x_2(s)), \label{eq:mat_law2}\\
u_3(s) & = g_3(x_3(s)). \label{eq:mat_law3}
\end{align}

\begin{sloppypar}
Let the first two axes in the body-fixed frame $a_1(s)$ and $a_2(s)$ correspond to the principal bending axes and the third axis $a_3(s)$ correspond to the principal torsion axis. Now, if the applied shear force is acting along axis $a_1$, then the bio-filament bends in-plane about axis $a_2$. Thus the first and third components of $\bfk(s)$ and $\bfq(s)$ along with the second component of $\bff(s)$ are zero. Thus the rod model reduces to the one-dimensional rod model equations
\begin{align}
\frac{dq_2}{ds} & = -{\rm f_1}(s), \label{eq:cant1} \\
\frac{df_1}{ds} & = -{\rm f_3}(s)\kappa_2(s), \label{eq:cant2}\\
\frac{df_3}{ds} & = {\rm f_1}(s)\kappa_2(s), \label{eq:cant3}
\end{align}
with the constitutive-law equation (\ref{eq:mat_law2}). For this simplified case, defining the state vector as $x^{1D}(s) = \matl{ccc} q_2(s) & f_1(s) & f_3(s) \matr^\rmT \in \BBR^3$ and the input as $u^{1D}(s) = \kappa_2(s) \in \BBR$, the one-dimensional rod model equations can be written in the state-space form as
\begin{align}
\frac{d}{ds}x^{1D}(s) &
 = \matl{c} -x^{1D}_2(s)\\ -x^{1D}_3(s)u^{1D}(s) \\ x^{1D}_2(s)u^{1D}(s)\matr. \label{eq:ss_model1D}
\end{align}
\end{sloppypar}

To illustrate the estimation procedure, we choose $g_2(\cdot)$ to be both an arctangent function and saturation function and set the loading at the free end to be $x(0) = \matl{ccc} 1 & 2 & 0 \matr$, where all numbers are dimensionless.



\begin{figure}
\begin{minipage}[t]{3.5in}
 \centering
 \includegraphics[scale=0.5]{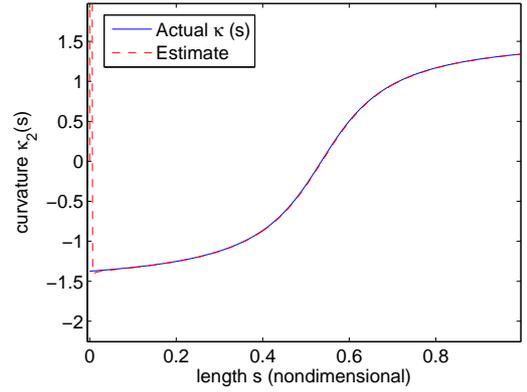}
 \parbox{3.5in}{\caption{Actual and estimated deformation $\bfk_2(s)$ for a cantilever DNA molecule with in-plane deformations. }
 \label{fig:kappa}}
 \end{minipage}
\end{figure}

Using measurements of $\bff(s)$ and $\bfq(s)$, in step 1, we use the UUMV filter with $s$ as the independent variable, to estimate $\bfk_2(s)$. Figure \ref{fig:kappa} shows estimate of $\kappa_2(s)$ using the UUMV filter. Note that the unknown loading conditions at the free end (unknown initial conditions) have limited detrimental effect on the estimate.


\begin{figure}[h!]
\begin{minipage}[t]{3.5in}
 \centering
 \includegraphics[scale=0.5]{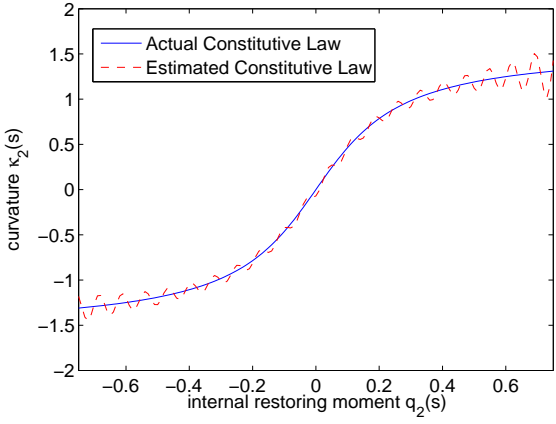}
 \parbox{3.5in}{\caption{Actual and estimated arctangent constitutive law for a DNA Molecule with the in-plane cantilever deformation.}
 \label{fig:atan}}
 \end{minipage}
  \end{figure}

\begin{figure}[h!]
 \begin{minipage}[t]{3.5in}
 \centering
 \includegraphics[scale=0.5]{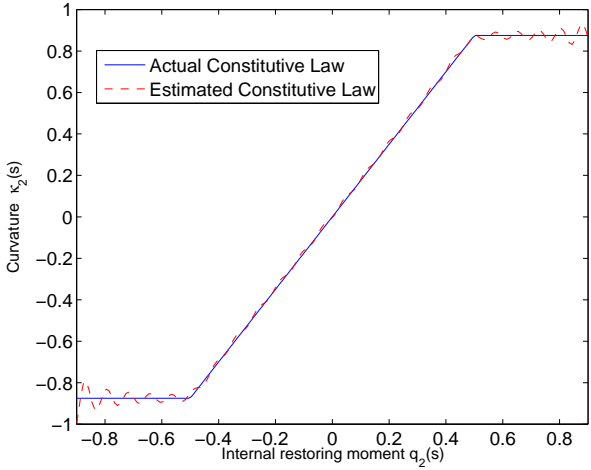}
 \parbox{3.5in}{\caption{Actual and estimated saturation constitutive law for a cantilever DNA molecule with in-plane deformation.}\label{fig:sat}}
 \end{minipage}
  \end{figure}

Once $\bfk_2(s)$ is estimated, in step 2, the function $g_2(\cdot)$ is estimated by representing the unknown function as an expansion of sinusoidal basis functions and using a standard least-squares to fit the unknown coefficients of the basis function expansion as in (\ref{eq:ls}). Figure \ref{fig:atan} shows the actual and estimate of the constitutive law $g_2(\cdot)$, when an arctangent function is used for simulations. Figure \ref{fig:sat} shows the actual and estimate of the constitutive law $g_2(\cdot)$, when a saturation function is used for simulations.

Finally, we note that by running three separate experiments with excitation along one principal axis at a time, we can use the same procedure discussed above to estimate all three constitutive laws (\ref{eq:mat_law1}) - (\ref{eq:mat_law3}), and thus the complete nonlinear constitutive law. The estimated constitutive law can then be used to predict deformations for any general loading configuration.


\section{Decoupled Three-dimensional Problem}

As discussed in the previous section, when the constitutive law is decoupled, the complete constitutive law can be estimated from three separate experiments focussing on one axis at a time. In this section, we consider the estimation of the complete decoupled constitutive law from a single experiment by choosing the loading conditions at the free end appropriately.

For step 1 of the two-step estimation technique, by using complete state-space equations (\ref{eq:ss_model}), we use the UUMV filter described in the appendix to estimate the unknown three-dimensional curvature vector $u(s) = \bfk(s).$ By using the estimated curvature vector $\hat \bfk(s)$ and the known internal force vector $\bfq(s)$, we use standard least squares fitting to estimate the coefficients of the basis function approximation (\ref{eq:ls}). 

\begin{figure}[h!]
 \begin{minipage}[t]{3.5in}
 \centering
 \includegraphics[scale=0.5]{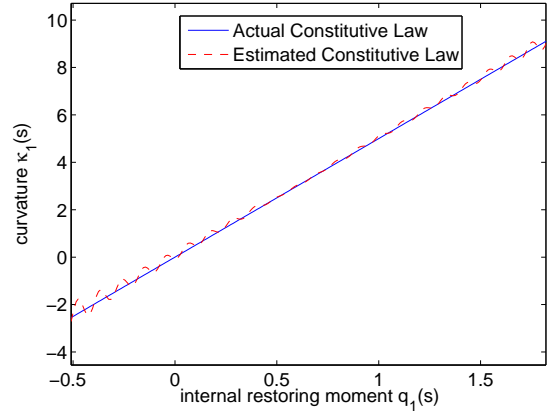}
 \parbox{3.5in}{\caption{Actual and estimated linear constitutive law for a cantilever DNA molecule with three-dimensional deformation. }\label{fig:TD_mat_law1}}
 \end{minipage}
\end{figure}

\begin{figure}[h!]
   \begin{minipage}[t]{3.5in}
 \centering
 \includegraphics[scale=0.5]{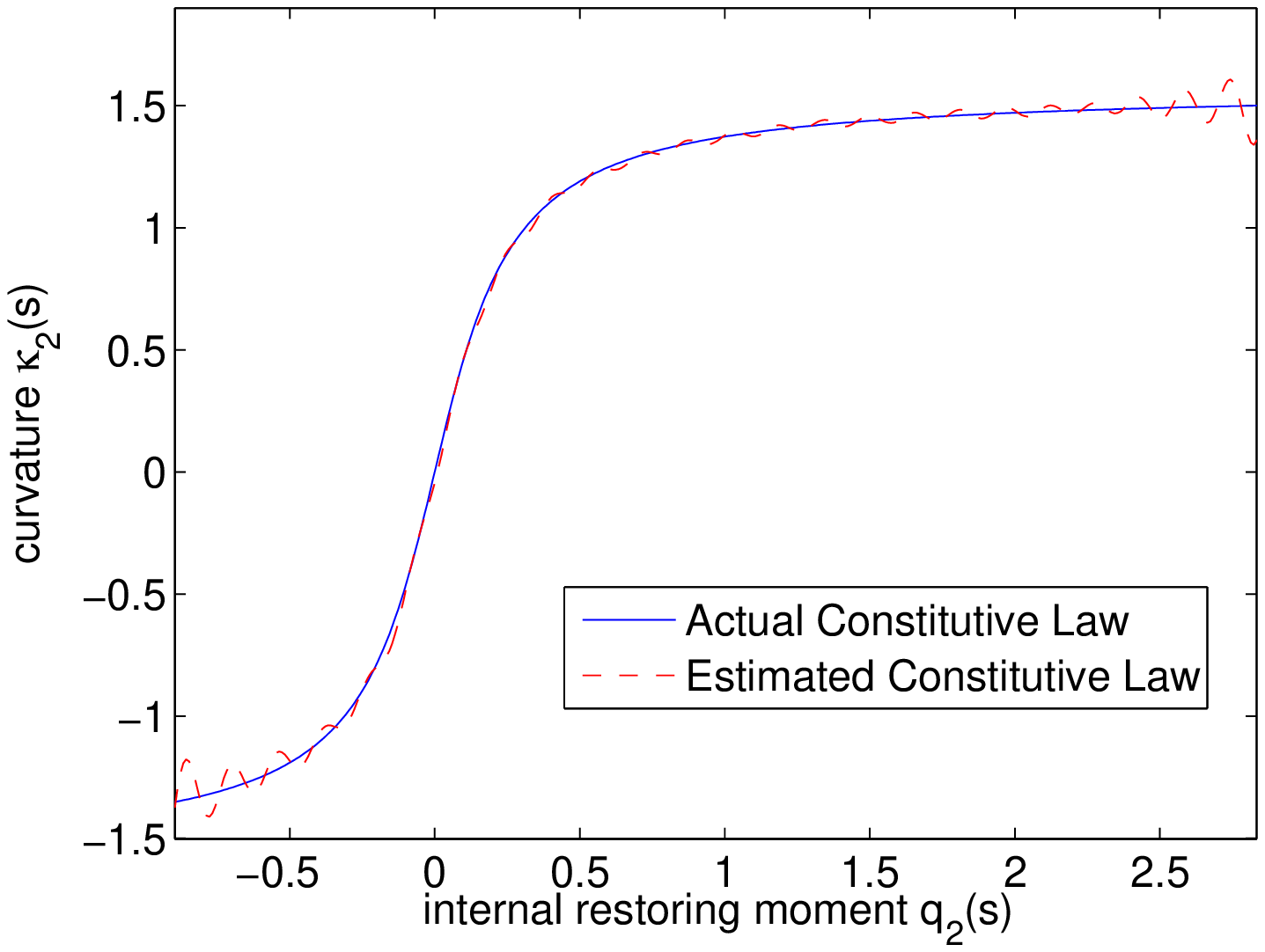}
 \parbox{3.5in}{\caption{Actual and estimated arctangent constitutive law for a cantilever DNA molecule with three-dimensional deformation.}\label{fig:TD_mat_law2}}
 \end{minipage}
\end{figure}

\begin{figure}[h!]
   \begin{minipage}[t]{3.5in}
 \centering
 \includegraphics[scale=0.5]{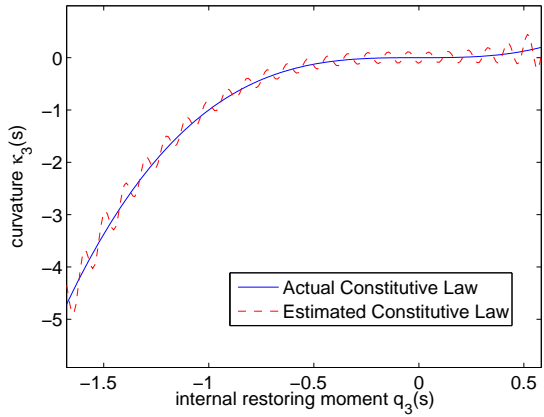}
 \parbox{3.5in}{\caption{Actual and estimated arctangent constitutive law for a cantilever DNA molecule with three-dimensional deformation.}\label{fig:TD_mat_law3}}
 \end{minipage}
\end{figure}

\begin{sloppypar}
The loading condition at the free end is chosen to be $x(0) = \matl{cccccc} 2 & -1 & 0 & -1 & -1 & -5 \matr$, while the three components of the decoupled constitutive law are chosen to be a linear function and two arctangent functions, respectively. Figures \ref{fig:TD_mat_law1}, \ref{fig:TD_mat_law2}, and \ref{fig:TD_mat_law3} show estimates of the three components of the constitutive law obtained as described above.
\end{sloppypar}

Note that the unknown-input matrix used in the UUMV filter is state-dependent and thus varying with the independent variable $s$
\begin{align}
B(s) = \matl{ccc} 0 & \rmq_3(s) & -\rmq_2(s) \\ - \rmq_3(s) & 0 & \rmq_1(s) \\ \rmq_2(s) & -\rmq_1(s) & 0 \\ 0 & \rmf_3(s) & -\rmf_2(s) \\ - \rmf_3(s) & 0 & \rmf_1(s) \\ \rmf_2(s) & -\rmf_1(s) & 0 \matr.
\end{align}
In the implementation of the UUMV filter, the state estimates are used to construct this state-dependent $B(s)$ matrix.

\section{Coupled Three-dimensional Problem}

In this section, we relax assumption A1. That is, we let the constitutive law be coupled, and assume a more general form of constitutive law written as
\begin{align}
\kappa_1(s) & = g_1(\bfq(s),\bff(s)), \label{eq:coupled1}\\
\kappa_2(s) & = g_2(\bfq(s),\bff(s)), \label{eq:coupled2}\\
\kappa_3(s) & = g_3(\bfq(s),\bff(s)). \label{eq:coupled3}
\end{align}

In principle, the coupled constitutive law does not affect the first step in which the unknown curvature vector $u(s)=\bfk(s)$ is estimated using the UUMV filter. However, the second step in which the estimated input $\hat \bfk(s)$ and the known state are used to estimate a coupled functional relationship of the form (\ref{eq:coupled1}) - (\ref{eq:coupled3}) using a basis-function fit (\ref{eq:ls}), differs in two ways. First, to estimate a coupled relationship,  multivariable basis functions have to be chosen for the basis-function expansion in (\ref{eq:ls_fit}). Here, we choose thin-plate-spline radial basis functions that have the form
\begin{align}
\phi_i(x(s)) = \| x(s)-c_i \|^2 \log \|x(s) - c_i\|,
\end{align}
where $c_i$ are centers of the radial basis functions and are chosen by the user. Second, from a practical point of view, a single experiment only provides data for a single curve on a multi-dimensional constitutive-law surface. Thus a single experiment does not provide enough data to estimate the coupled constitutive law. To deal with this issue, we used an ensemble of 50 experiments with different loading conditions, and thus generating data that represents 50 different curves on the multi-dimensional constitutive-law surface.

By using an ensemble of experiments, we use the UUMV filter to generate estimates of the curvature vector $\bfk(s)$ for all 50 experiments. Using this ensemble of estimates of $\bfk(s)$ and known $\bfq(s)$ and $\bff(s)$, a single least-squares problem for estimating the coefficients of the basis function expansion (\ref{eq:ls}) is solved. Note that since the UUMV filter is robust to unknown initial conditions, the loading conditions for the 50 different experiments need not be known.

\begin{figure}[h!]
   \begin{minipage}[t]{3.5in}
 \centering
 \includegraphics[scale=0.5]{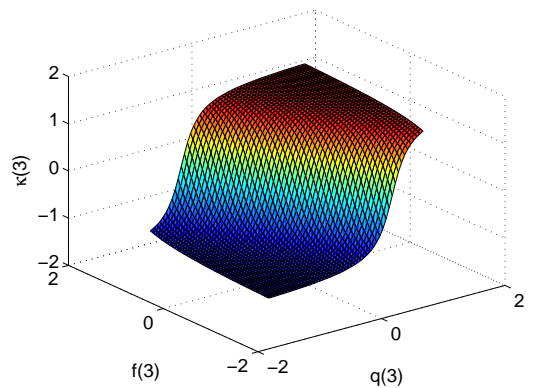}
 \parbox{3.5in}{\caption{Actual coupled nonlinear constitutive law for a cantilever DNA molecule.}\label{fig:coupled_actual}}
 \end{minipage}
  \end{figure}

\begin{figure}[h!]
   \begin{minipage}[t]{3.5in}
 \centering
 \includegraphics[scale=0.5]{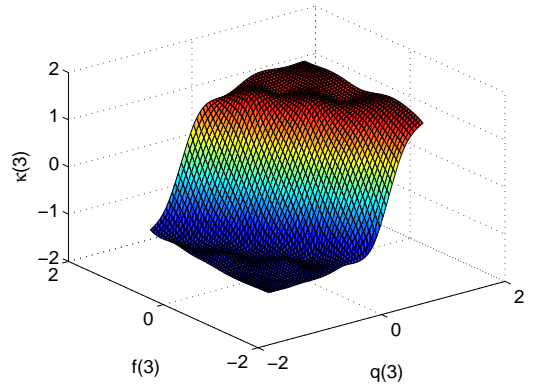}
 \parbox{3.5in}{\caption{Estimate of the coupled nonlinear constitutive law for a cantilever DNA molecule with three-dimensional deformation.}\label{fig:coupled_estimate}}
 \end{minipage}
\end{figure}

\begin{sloppypar}
For demonstrating the above technique, we use the 2-D coupled constitutive law $\kappa_3(s) = \tan^{-1}(5f_3(s)q_3(s))$. Figure \ref{fig:coupled_actual} shows the second component of the actual constitutive law, while Figure \ref{fig:coupled_estimate} shows it's estimate using the above-described technique.
\end{sloppypar}
%

\begin{figure}[h!]
   \begin{minipage}[t]{3.5in}
 \centering
 \includegraphics[scale=0.5]{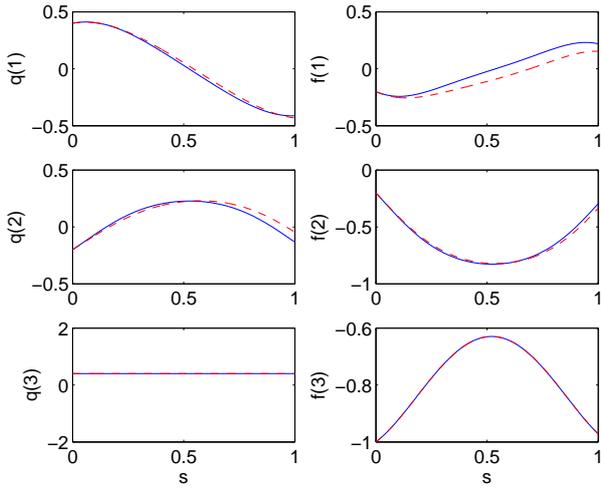}
 \parbox{3.5in}{\caption{Comparison of elastic rod model simulations with the true constitutive law and with the estimated constitutive law for an independent set of loading conditions.}\label{fig:validation_coupled}}
 \end{minipage}
\end{figure}

Finally, Figure \ref{fig:validation_coupled} shows a validation check using a comparison of the results of the elastic rod model simulations with the actual constitutive law and with the estimated constitutive law for a set of independent loading conditions that were not used to estimate the constitutive law.


\section{Robustness}

Next, we relax assumptions A2 and A3, and thus study the robustness of the algorithm under sparse and/or noisy measurements. For this analysis, we use the decoupled one-dimensional problem discussed in section \ref{oneDprob} for convenience. We note that the analysis provided here is also applicable to the decoupled three-dimensional problem and the coupled three-dimensional problem since the underlying algorithms are the same. To assess the quality of the estimate, we use the normalized mean square error (MSE) defined as
\begin{align}
{\rm MSE} = \frac{1}{N}\sum_{i=1}^N \left(\kappa(s_i) - \hat \kappa (s_i) \right)^2,
\end{align}
where $\hat \kappa$ represents the estimate of $\kappa,$ $s_i$ represents the measurement locations, and $N$ is the total number of interior measurement locations.

\begin{figure}[h!]
   \begin{minipage}[t]{3.5in}
 \centering
 \includegraphics[scale=0.35]{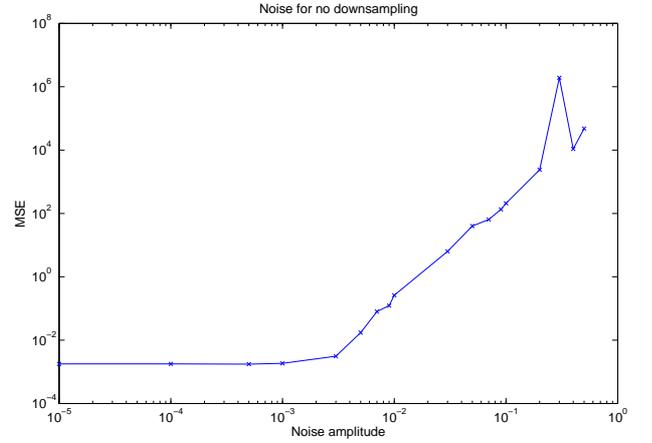}
 \parbox{3.5in}{\caption{MSE as a function of noise amplitude with no downsampling.}\label{fig:MSE_noise_nodownsampling}}
 \end{minipage}
\end{figure}

\begin{figure}[h!]
 \begin{minipage}[t]{3.5in}
 \centering
 \includegraphics[scale=0.35]{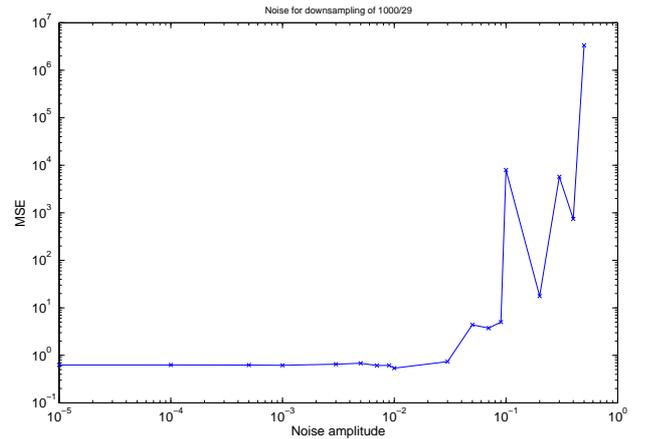}
 \parbox{3.5in}{\caption{MSE as a function of noise amplitude with using 30 points out of 1000.}\label{fig:MSE_noise_downsampling29}}
 \end{minipage}
\end{figure}

Figure \ref{fig:MSE_noise_nodownsampling} represent the MSE as a function of increasing noise amplitude $v_k$. A total of 1000 interior measurement points are used for each simulation in the plot. On the other hand, Figure \ref{fig:MSE_noise_downsampling29} represent the MSE as a function of increasing noise amplitude with only 30 interior measurement points. As expected, the MSE for low noise amplitudes in Figure \ref{fig:MSE_noise_nodownsampling} is lower than in Figure \ref{fig:MSE_noise_downsampling29}. However, the differences at high noise amplitude are smaller, indicating the beyond a noise amplitude of $10^{-2},$ the detrimental effect of the noise is dominant over the detrimental effect due to sparse data. Furthermore, in both plots, the MSE is flat for a significant portion of the low noise amplitude range. Suggesting that the algorithms are insensitive to noise at these low amplitudes.

\begin{figure}[h!]
   \begin{minipage}[t]{3.5in}
 \centering
 \includegraphics[scale=0.35]{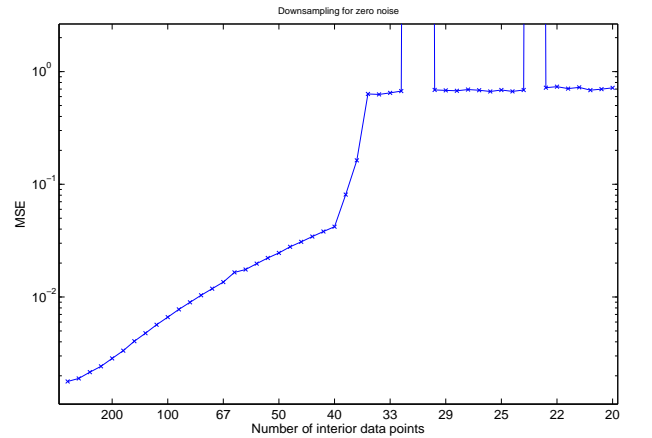}
 \parbox{3.5in}{\caption{MSE as a function of downsampling for zero noise.}\label{fig:MSE_downsampling_zeronoise}}
 \end{minipage}
\end{figure}

\begin{figure}[h!]
   \begin{minipage}[t]{3.5in}
 \centering
 \includegraphics[scale=0.35]{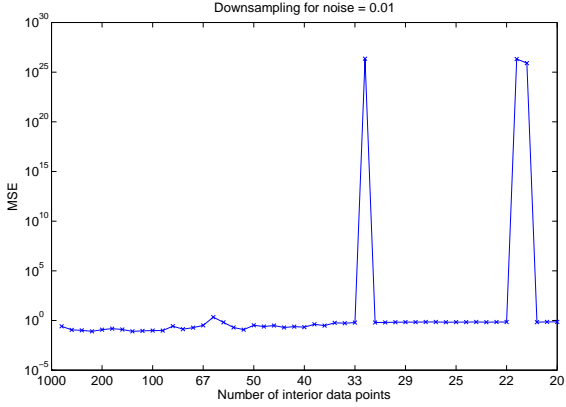}
 \parbox{3.5in}{\caption{MSE as a function of downsampling for noise amplitude 0.01.}\label{fig:MSE_downsampling}}
 \end{minipage}
\end{figure}

Figure \ref{fig:MSE_downsampling_zeronoise} represent the MSE as a function of decreasing interior measurement points. The measurements were assumed to be noise-free. This plot suggests that there exists a threshold beyond which, sparser data do not deteriorate the estimates any further. Finally, Figure \ref{fig:MSE_downsampling_zeronoise} also seems to confirm the belief that more data is always beneficial. Figure \ref{fig:MSE_downsampling} represent the MSE as a function of decreasing interior measurement points, a gaussian white noise of standard deviation 0.01 was used to corrupt all measured variables. Again, the flat nature of this plot confirms that a noise amplitude of 0.01 dominates the effect due to sparse measurements.

\begin{figure}[h!]
   \begin{minipage}[t]{3.5in}
 \centering
 \includegraphics[scale=0.35]{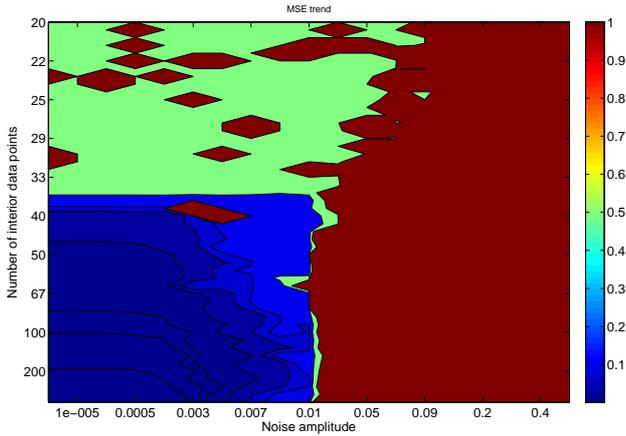}
 \parbox{3.5in}{\caption{MSE as a function of noise and downsampling.}\label{fig:MSE_trend}}
 \end{minipage}
\end{figure}

Finally, Figure \ref{fig:MSE_trend} shows MSE as a function of increasing noise amplitude and decreasing number of interior measurement points. The effect due to noise amplitude and sparse measurements seem to be clearly decoupled with acceptable accuracy with a region with noise standard deviation 0.01 or less and interior measurement points of 30 or more.

\section{Partial State Measurements}

Finally, we consider the cases in which the output measurements $y(s)$ are a subset of the states $x(s)$. As noted earlier, in these cases, the measurement equation becomes
\begin{align}
y(s) = Cx(s).
\end{align}

Since, we need $l \geq m$ for (\ref{eq:constraint1}) to be satisfied, we examine the most difficult scenario $l = m = 3.$ We consider the following representative cases for $C$ for $l = 3$.
\begin{align}
C & = \matl{cc} I_3 & 0_{3 \times 3} \matr, \label{eq:C1} \\
C & = \matl{cc} 0_{3 \times 3} & I_3 \matr,\label{eq:C2} \\
C & = \matl{cccccc} 0 & 1 & 0 & 0 & 0 & 0 \\ 0 & 0 & 0 & 1 & 0 & 0\\ 0 & 0 & 0 & 0 & 0 & 1 \matr. \label{eq:C3}
\end{align}

\begin{sloppypar}
As discussed in Section \ref{sec:IR}, a necessary condition for unbiased input reconstruction is that for $s$ such that rank$(B(s)) = m,$ rank$(CB(s)) =$ dim$ (u(s)) = 3$. It follows that $C$ in (\ref{eq:C1}) and (\ref{eq:C2}) do not satisfy this condition, while $C$ in (\ref{eq:C3}) satisfies this necessary condition. Therefore, it follows that measurements of either $f(s)$ or $q(s)$ alone does not help determine the constitutive law. Thus experimentalists will have to design experiments in such a way that a combination of components of $f(s)$ and $q(s)$ are measured.
\end{sloppypar}

\section{Discussion}
\label{discussion}

Although the results in this paper show promise, several practical and theoretical issues remain. Ongoing and future work includes estimating an implicit constitutive law that is dependent on $s$. For both a implicit constitutive law and $s$-dependent constitutive law, the input reconstruction step to estimate $\bfk(s)$ remains unchanged. However, once estimates $\hat{\bfk}(s)$ are obtained, a more sophisticated  function approximation technique must be used for the second step. Moreover, for an implicit constitutive law, the implementation of the final elastic rod model equations with the estimated constitutive law must be using a DAE solver.

Furthermore, in short length scales, the disturbances experienced by DNA molecules due to thermal fluctuations of the surrounding aqueous medium and hydrodynamic forces often overwhelm deterministic forces and loading conditions applied to DNA molecules. Thus, the estimation algorithms must be robust for state disturbance $w_k$ of magnitude larger than the deterministic deformations. 

Finally, measurements in current DNA experiments are bulk properties of the molecules such as the elastic strain energy which can be expressed as
\begin{align}
U = \int_0^L \int_{\bfk_0(s)}^{\bfk(s)} \bfq(s)\cdot dk \ ds. \label{eq:measurement}
\end{align}
Note that (\ref{eq:measurement}) is a function of the states at several values of the independent variable $s$ and does not readily fit into the Kalman filtering and UUMV filtering framework. Both these significant challenges will be addressed as part of future work.

\section{Conclusion}
We developed a two-step technique to use elastic rod model equations in combination with limited experimental measurements or high-fidelity molecular dynamics simulation data to estimate the nonlinear constitutive law governing DNA molecules. We first cast the elastic rod model equations in state-space form and expressed the effect of the unknown constitutive law as an unknown input to the system. Then, in step 1, we used input reconstruction techniques to estimate unmeasured states and unknown inputs of rod model equations. In step 2, estimates from step 1 were used in least-squares function fitting to approximate the constitutive law. Various simplification and scenarios with decoupled constitutive laws and coupled constitutive laws were discussed. We investigated the robustness of the two-step technique through simulations and made several observations. We finished with some concluding discussion about future work.




\footnotesize
\bibliographystyle{plain}
\bibliography{jrnls_short,bishop,asme2e,DNA_refs}



\normalsize

\section*{Appendix}

\subsection*{State Estimation for Nonlinear Systems}
\label{sec:ss_nl}

\begin{sloppypar}
Consider the nonlinear stochastic discrete-time dynamic system
\begin{eqnarray}
\label{nlmodel1} {x}_{k+1}  &=&  \psi\left({x}_{k},{u}_{k},{w}_{k}\right), \\
\label{nlmodel2} {y}_k  &=&  h\left({x}_{k}\right) + {v}_{k},
\end{eqnarray}
where $\psi: \bbR^n \times \bbR^m \times \bbR^{q} \rightarrow
\bbR^n$ and $h: \bbR^n \rightarrow \bbR^l$ are, respectively,
the process and observation models. The optimal solution to the  
state-estimation problem is complicated \cite{daum/2005} by the fact that, for nonlinear
systems, $\rho({x}_k|({y}_{1},\ldots,{y}_{k}))$ is not completely
characterized by its first and second-order moments. We thus use
an approximation based on the classical Kalman filter to provide a
suboptimal solution to the nonlinear case.
\end{sloppypar}

\subsection*{Unscented Kalman Filter}

First, for nonlinear systems, we consider the unscented Kalman
filter (UKF) \cite{julier_uhlmann/2004} to provide a suboptimal
solution to the state-estimation problem. Instead of analytically
linearizing the nonlinear state-space model and using linear filter 
equations, UKF employs the unscented
transform (UT) \cite{julier_etal/2000}, which approximates the
posterior mean $\bar{y} \in \bbR^l$ and covariance $P^{yy} \in
\bbR^{l \times l}$ of a random vector $y$ obtained from the
nonlinear transformation $y = h(x)$, where ${x}$ is a prior random
vector whose mean $\bar{x} \in \bbR^n$ and covariance $P^{xx} \in
\bbR^{n \times n}$ are assumed known. UT yields the actual mean
$\bar{y}$ and the actual covariance $P^{yy}$ if $h = h_{1} + h_{2}$,
where $h_{1}$ is  linear and $h_{2}$ is quadratic
\cite{julier_etal/2000}. Otherwise, $\hat{y}_{k}$ is a {\em pseudo
mean} and $P^{yy}$ is a {\em pseudo covariance}.

UT is based on a set of deterministically chosen vectors known as
sigma points.
To capture the mean $\bar{x}_{k}^{\rm a}$ of the augmented prior
state vector
\begin{equation}
\label{x_aug} {x}_{k}^{\rm a} ~\triangleq~ \left[
\begin{array}{c} {x}_{k} \\ w_{k} \end{array} \right],
\end{equation}
where ${x}_{k}^{\rm a} \in \bbR^{n_{\rm a}}$ and $n_{\rm a}
\triangleq n +  q$, as well as the augmented prior error
covariance
\begin{eqnarray}
\label{Pa} P^{xx{\rm a}}_{k} \triangleq \left[
\begin{array}{c c}
P^{xx}_{k+1|k} & 0_{n \times  q} \\
0_{ q \times n} & Q_{k}\\
\end{array}
\right],
\end{eqnarray}
the sigma-point matrix ${\cal X}_{k} \in \bbR^{n_{\rm a} \times
(2n_{\rm a} + 1)}$ is chosen as
\begin{eqnarray}
\centering \label{sigmapoints} \nn \left\{
\begin{array}{ll}
{\rm col}_0({\cal X}_{k}) & \triangleq  \hat{x}_{k}^{\rm a}, \\
{\rm col}_i({\cal X}_{k}) & \triangleq  \hat{x}_{k}^{\rm a} \\
&  ~+ ~ \sqrt{(n_{\rm a}+\lambda)}\,{\rm col}_i\left[{\left(P_{k}^{xx{\rm a}}\right)}^{1/2} \right], \\
& \quad i = 1, \ldots, n_{\rm a}, \\
{\rm col}_{i+n_{\rm a}}({\cal X}_{k})  & \triangleq
\hat{x}_{k}^{\rm a} \\
& ~-~ \sqrt{(n_{\rm a}+\lambda)}\,{\rm
col}_i\left[{\left(P_{k}^{xx{\rm a}}\right)}^{1/2} \right],
\\ & \quad i = 1, \ldots, n_{\rm a},
\end{array}
\right.
\end{eqnarray}
with weights
\begin{eqnarray} \label{weights} \nn
\left\{
\begin{array}{lcl}
 \gamma_0^{(m)} & \triangleq & {\displaystyle \frac{\lambda}{n_{\rm a} + \lambda}},  \\
\gamma_0^{(c)} & \triangleq & {\displaystyle \frac{\lambda}{n_{\rm a} + \lambda} + 1 - \alpha^2 + \beta},  \\
 \gamma_i^{(m)} & \triangleq & \gamma_i^{(c)} ~\triangleq~~ \gamma_{i + n_{\rm a}}^{(m)} ~\triangleq~~
\gamma_{i + n_{\rm a}}^{(c)} ~\triangleq~~ {\displaystyle
\frac{1}{2(n_{\rm a} + \lambda)}}, \\
& & \quad i = 1, \ldots, n_{\rm a},
\end{array}
\right.
\end{eqnarray}
where ${\rm col}_i\left[(\cdot)^{1/2}\right]$ is the $i$th column
of the Cholesky square root, $ 0 < \alpha \le 1$, $\beta \ge 0$,
$\theta \ge 0$, and $\lambda \triangleq \alpha^2(\theta + n_{\rm
a}) - n_{\rm a}$. We set $\alpha=1$ and  $\theta=0$
\cite{haykin/2001} such that $\lambda = 0$
\cite{julier_uhlmann/2004} and set $\beta = 2$ \cite{haykin/2001}.
Alternative schemes for choosing sigma points are given in
\cite{julier_uhlmann/2004}.

The UKF {\em forecast} equations are given by
\newpage
\begin{eqnarray}
 {\cal X}_{k} & = & \left[\hat{x}_{k}^{\rm a} \quad \hat{x}_{k}^{\rm a} 1_{1 \times n_{\rm a}} + \sqrt{(n_{\rm a}+\lambda)}{\left(P_{k}^{xx{\rm a}}\right)}^{1/2} \quad \hat{x}_{k}^{\rm a} 1_{1 \times n_{\rm a}} - \sqrt{(n_{\rm a}+\lambda)}{\left(P_{k}^{xx{\rm a}}\right)}^{1/2} \right], \nonumber \\ \label{ukf1a}
\end{eqnarray}
\begin{align}
\label{ukf1a_1} {\rm col}_i({\cal X}_{k+1|k}^{x}) & = \psi( {\rm col}_i({\cal X}_{k}^{x}),~ {u}_{k},~{\rm col}_i({\cal X}_{k}^{w})), \quad  i = 0, \ldots, 2n_{\rm a}, \\
\label{ukf1a_2} \hat{x}_{k+1|k} & =  \sum_{i=0}^{2n_{\rm a}} \gamma_i^{(m)} {\rm col}_i({\cal X}_{k+1|k}^{x}),  \\
\label{ukf1b} P_{k+1|k}^{xx} & = \sum_{i=0}^{2n_{\rm a}} \gamma_i^{(c)} [{\rm col}_i({\cal X}_{k+1|k}^{x}) -\hat{x}_{k+1|k} ][{\rm col}_i({\cal X}_{k+1|k}^{x}) -\hat{x}_{k+1|k}]^{\T},\\
\label{ukf1c_1} {\rm col}_i({\cal Y}_{k+1|k}) & = h( {\rm col}_i({\cal X}_{k | k-1}^x)), \quad  i = 0, \ldots, 2n_{\rm a}, \\
\label{ukf1c} \hat{y}_{k+1|k} & = \sum_{i=0}^{2n_{\rm a}} {\gamma_i^{(m)}} {\rm col}_i({\cal Y}_{k+1|k}), \\
\label{ukf1d} P_{k+1|k}^{yy} & = \sum_{i=0}^{2n_{\rm a}} {\gamma_i^{(c)}} [{\rm col}_i({\cal Y}_{k+1|k})-\hat{y}_{k+1|k} ][{\rm col}_i({\cal Y}_{k+1|k})-\hat{y}_{k+1|k}]^{\T} + R_k,\\
\label{ukf1e} P_{k+1|k}^{xy} & =  \sum_{i=0}^{2n_{\rm a}}
{\gamma_i^{(c)}} [{\rm col}_i({\cal X}_{k+1|k}^x) -\hat{x}_{k+1|k}
][{\rm col}_i({\cal Y}_{k+1|k})-\hat{y}_{k+1|k}]^{\T},
\end{align}
where $\left[ \begin{array}{c} {\cal X}_{k}^{x} \\ {\cal
X}_{k}^{w} \end{array} \right] \triangleq {\cal X}_{k}$,
${\cal X}_{k}^{x} \in \bbR^{n \times (2n_{\rm a} + 1)}$, and
${\cal X}_{k}^{w} \in \bbR^{ q \times (2n_{\rm a} + 1)}$.

\subsection*{Unbiased Minimum-variance Unscented Filter}

Next, for nonlinear systems with unknown inputs, we consider an
extension of the UKF along the lines of the linear unbiased minimum-variance filter. Thus,
to obtain the pseudo mean and the pseudo error covariances we use
the unscented transform, and to estimate the states and unknown
inputs, we use the expressions derived for the unbiased minimum-variance filter. Thus, the
{\it forecast} equations for the unbiased minimum-variance unscented
(UMVU) filter are given by (\ref{ukf1a}) - (\ref{ukf1e}). The {\it
data-assimilation} equations for the UMVU filter are given by
\begin{align}
\hat{x}_{k+1|k+1}  &= \hat{x}_{k+1|k} + + L_{k+1}(y_{k+1} - h(\hat{x}_{k+1|k})), \\
\hat{x}_{k+1|k}  &=  \psi\left(\hat{x}_{k|k},0,0\right).
\end{align}

\end{document}